\documentclass[aip, apl, showpacs, reprint, preprintnumbers,amsmath,amssymb,superscriptaddress,floatfix]{revtex4-1}

\usepackage[]{graphicx}      
\usepackage{bm}             	
\usepackage{times}			
\usepackage{tabularx}
\usepackage{color}
\usepackage{dcolumn}		
\usepackage{amsmath}
\usepackage{amssymb}
\usepackage{amsfonts}
\usepackage{times}
\usepackage{tabularx}
\usepackage{xcolor,colortbl}

\usepackage{amsmath}

\makeatletter
\newcommand\xleftrightarrow[2][]{%
  \ext@arrow 9999{\longleftrightarrowfill@}{#1}{#2}}
\newcommand\longleftrightarrowfill@{%
  \arrowfill@\leftarrow\relbar\rightarrow}
\makeatother

\begin{document}
\title{Annealing stability of magnetic tunnel junctions based on dual MgO free layers and [Co/Ni] based thin synthetic antiferromagnet fixed system}
\author{T. Devolder}
\email{thibaut.devolder@u-psud.fr}
\affiliation{Centre de Nanosciences et de Nanotechnologies, CNRS, Univ. Paris-Sud, Universit\'e Paris-Saclay, C2N-Orsay, 91405 Orsay cedex, France}
\author{S. Couet}
\affiliation{imec, Kapeldreef 75, 3001 Heverlee, Belgium}
\author{J. Swerts}
\affiliation{imec, Kapeldreef 75, 3001 Heverlee, Belgium}
\author{E. Liu}
\affiliation{imec, Kapeldreef 75, 3001 Heverlee, Belgium}
\affiliation{Department of Electrical Engineering (ESAT), KU Leuven, 3001 Leuven, Belgium}
\author{T. Lin}
\author{S. Mertens}
\author{A. Furnemont}
\author{G. Kar}
\affiliation{imec, Kapeldreef 75, 3001 Heverlee, Belgium}

\date{\today}                                           
%
%
\begin{abstract}
We study the annealing stability of bottom-pinned perpendicularly magnetized magnetic tunnel junctions based on dual MgO free layers and thin fixed systems comprising a hard [Co/Ni] multilayer antiferromagnetically coupled to thin a Co reference layer and a FeCoB polarizing layer. Using conventional magnetometry and advanced broadband ferromagnetic resonance, we identify the properties of each sub-unit of the magnetic tunnel junction and demonstrate that this material option can ensure a satisfactory resilience to the 400$^\circ$C thermal annealing needed in solid-state magnetic memory applications. The dual MgO free layer possesses an anneal-robust 0.4 T effective anisotropy and suffers only a minor increase of its Gilbert damping from 0.007 to 0.010 for the toughest annealing conditions. Within the fixed system, the ferro-coupler and texture-breaking TaFeCoB layer keeps an interlayer exchange above 0.8 mJ/m$^2$, while the Ru antiferrocoupler layer within the synthetic antiferromagnet maintains a coupling above -0.5 mJ/m$^2$. These two strong couplings maintain the overall functionality of the tunnel junction upon the toughest annealing despite the gradual degradation of the thin Co layer anisotropy that may reduce the operation margin in spin torque memory applications. Based on these findings, we propose further optimization routes for the next generation magnetic tunnel junctions.
\end{abstract}

\maketitle

%
%

\section{Introduction}
The spin-transfer-torque (STT) switching of a nanomagnet that shows perpendicular magnetic anisotropy is the basic phenomenon harnessed in 
advanced magnetic random access memories. This technology is based on magnetic tunnel junctions (MTJ)  that rely on complex stacks that require careful optimization to ensure manufacturability and operability.
Thanks to their low Gilbert damping and their high performing magneto-transport properties \cite{ikeda_perpendicular-anisotropy_2010} the so-called single MgO  free layer of composition Ta/FeCoB/MgO has become \textit{de facto} the standard storage layer. Unfortunately in single MgO free layers only one interface contributes to the total magneto-crystalline anisotropy. Besides, some diffusion of Ta into FeCoB happens at the high annealing temperature required for a CMOS back-end-of-line process, and this intermixing is detrimental \cite{devolder_irradiation-induced_2013} to the Gilbert damping.  Dual MgO (i.e. MgO/FeCoB/spacer/FeCoB/MgO) free layers are thus gradually introduced in the context of  spin-torque operated magnetic memory applications as they provides superior properties with the promise of better resilience to annealing \cite{sato_properties_2014, kim_ultrathin_2015}.

On the other side of the tunneling oxide, magnetically hard reference layers are needed. A synthetic antiferromagnet (SAF) system is systematically used to avoid generating stray fields that would perturb the free layer operation. Historically, two multilayers [Co/X]$_{\times N}$ and [Co/X]$_{\times M}$ with $\textrm{X} \in [\textrm{Ni,~Pt,~Pd}]$ separated by a Ru-based antiferrocoupler have been used \cite{worledge_spin_2011} to form a conventional thick SAF. Sadly, conventional SAF do not scale well. Indeed the stray field compensation requires the farthest multilayer to be much thicker than the one nearest to MgO, with a ratio $M/N$ that has to be increased substantially at low junction diameter \cite{gottwald_scalable_2015} with the consequent growth and etching challenges. Thin SAFs, i.e. with a single [Co/X] multilayer ($N=1$) were proposed \cite{gan_perpendicular_2014, tomczak_thin_2016} instead to maintain manufacturability at low junctions diameters.  

In this paper, we optimize and evaluate the annealing resistance of bottom pinned MTJs comprising a dual MgO free layer and a [Co/Ni]-based thin SAF in which the near MgO magnet of the SAF has been simplified to a solitary Co layer. Using broadband ferromagnetic resonance, we study the properties of each sub-unit of the MTJ and demonstrate that this material option can ensure a good resilience to 400$^\circ$C thermal annealing of the transport properties of the MTJ and the damping of the free layer, despite the gradual evolution of the thin Co layer whose anisotropy and magnetization degrades during the annealing.

%
\begin{figure}
\includegraphics[width=8.5 cm]{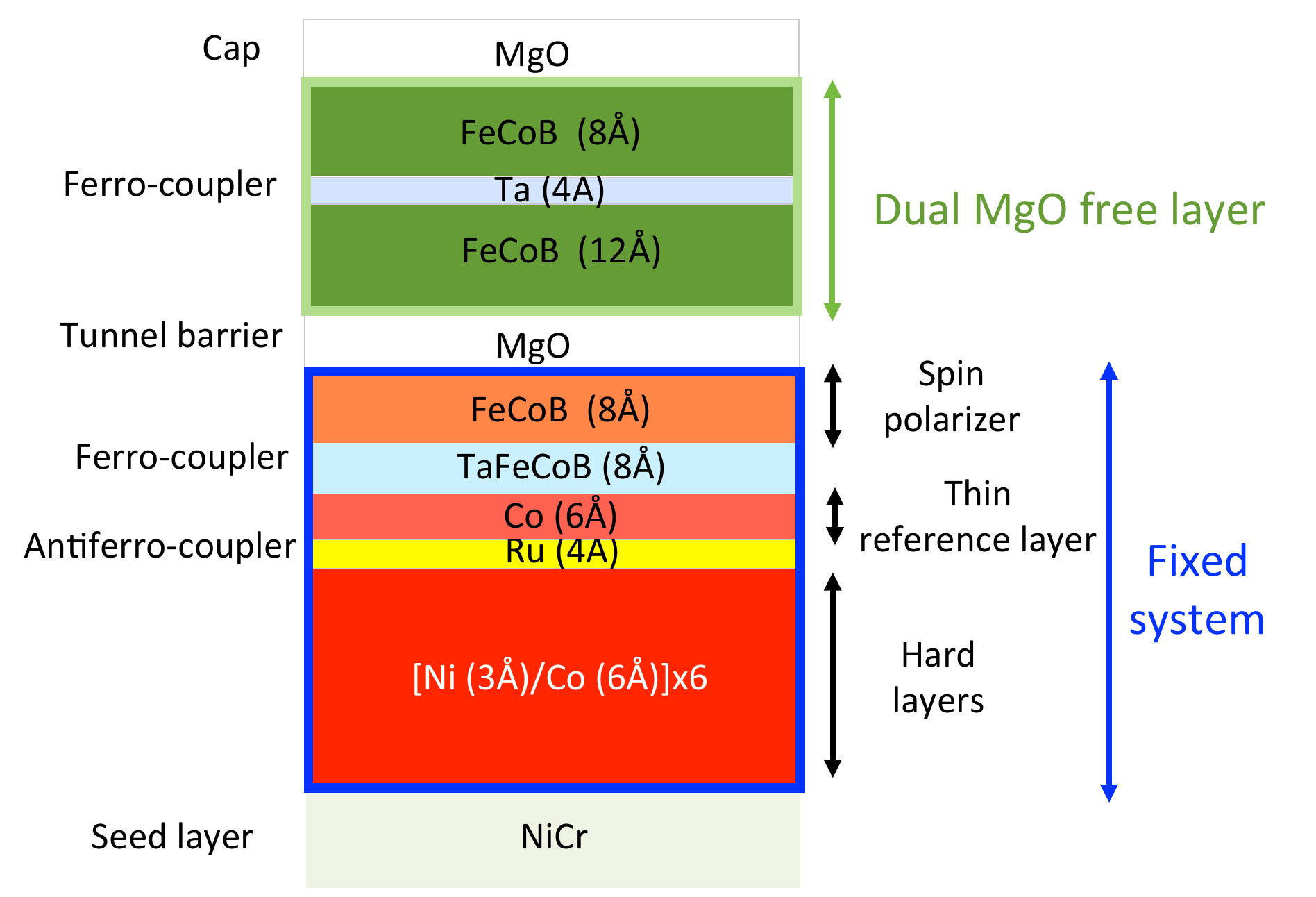}
\caption{(Color online). Sketch of the Magnetic Tunnel Junction with a dual MgO free layer and a [Co/Ni] multilayer based thin synthetic antiferromagnet fixed system. Note our layer labeling  conventions within the fixed system.}
\label{stack}
\end{figure}

%
\begin{figure}
\caption{(Color online). Properties of the dual MgO free layer. (a) to (d): Imaginary part of the free layer permeability in a field of 1 Tesla along the easy axis for the four annealing conditions. The lines are macrospin fits with the linewidth $\Delta f / (2f) $ as fitting parameter.  (e) Major hysteresis loop of the full tunnel junction after the 300$^\circ$C annealing step. (f) Ferromagnetic resonance frequency versus field curves for the four annealing conditions. The four curves overlap within the experimental accuracy. (g) Half linewidth versus FMR frequency for the four annealing conditions. The lines have slopes of 0.007 (black empty circles, 300$^\circ$C) and 0.01 (orange filled circles, 400$^\circ$C).}
\label{FL}
\end{figure}

\section{Sample and Methods}
We study bottom-pinned MTJs (Fig.~\ref{stack}) of the following composition: NiCr (seed) / [Ni(3)/Co(6)]$_{\times 6}$ / Ru(4) / Co(6, RL) / TaFeCoB(8)/ FeCoB /  MgO / Free layer / MgO / cap, where the Free layer is composed of Fe$_{60}$Co$_{20}$B$_{20}$ (12) / Ta / Fe$_{60}$Co$_{20}$B$_{20}$ (8), with the numbers being the thickness in \r{A}. Our layer naming convention is described in Fig.~\ref{stack}. The Ta spacer within the free layer was grown with a sacrificial Mg layer to avoid alloying of Ta and FeCoB during growth \cite{swerts_beol_2015}. The TaFeCoB spacer used between the spin polarizing layer and the reference layer was deposited by co-sputtering from two Fe$_{60}$Co$_{20}$B$_{20}$ and Ta targets, with the Ta sputtering rate set to achieve a content above 50\%. Note that the Ru thickness was set at the first (and in principle the strongest) maximum of antiferromagnetic coupling, in contrast to the common practice with conventional SAFs. This choice might be perceived as surprising as a too low Ru thickness endangers the overall resistance to annealing; however we will see that the use of this Ru thickness nominally optimizing the coupling was found necessary to compensate for the fact that the Ru coupling does not fully develop for ultrathin Co(6\r{A}) reference layers like ours (Fig.~\ref{stack}). Note also that the Ru antiferrocoupler is sandwiched between two Co layers as this is recommended to maximize the interlayer exchange coupling. All samples have been annealed at 300$^{\circ}$C for 30 minutes in a field of 1 Tesla, followed by a rapid thermal annealing of 10 minutes at $T_a=$ 350, 375 or 400$^{\circ}$C.
\begin{table}
  \centering
  \begin{tabular}{|c|c|c|c|c|c|c}
  \hline
$T_\textrm{anneal}$ & $\alpha_\textrm{~FL}$ & $\frac{1}{\gamma_0} \Delta f |_{H=0}$     & $\mu_0(H_{k}-M_S)$   & TMR & RA \\  
 ($^{\circ}$C) & ($\pm$ 0.001) &  ($\pm$ 1.5 mT) & ($\pm$ 5 mT) &  (\%)  & $(\Omega. \mu \textrm{m}^2)$ \\  \hline
300 & 0.007 &  8  &  410 &   111 & 9.0\\  
350 &  0.008 & 11  &  420 &   132 & 9.5\\   
375 & 0.009 &  11 &  405   & 146 & 9.8\\   
400 & 0.010 & 10  &  410  &  144 & 10.8\\  \hline
  \end{tabular}
\caption{Free layer and transport properties versus annealing for our MgO/Fe$_{60}$Co$_{20}$B$_{20}$/Ta/Fe$_{60}$Co$_{20}$B$_{20}$ free layer.}
\label{TMR}
\end{table}

We studied our samples by Vector Network Ferromagnetic resonance (VNA-FMR, \cite{bilzer_vector_2007}), vibrating sample magnetometry (VSM) and current-in-plane tunneling (CIPT). In VSM, the loop of the fixed system was deduced by subtracting the square and low coercivity minor loop of the free layer (not shown) from the major loop of the full MTJ [Fig.~\ref{FL}(e)]. As the free layer switching field slightly fluctuates from loop to loop, some fuzziness appears inevitably at the free layer coercivity in the so-calculated hysteresis loop of the fixed layer system (blue curves, Fig.~\ref{CoNi}). 
We then use VNA-FMR to measure the MTJ eigenexcitations in out-of-plane fields up to 2.5~T and 70~GHz. 
The frequency resolution of VNA-FMR is a convenient tool to identify selectively the properties of each subsystem. The canonical method was described in detail in ref.~\onlinecite{devolder_evolution_2016}. For instance the free layer FMR frequency versus field curve changes slope at the free layer coercivity, while the fixed system eigenmodes undergo frequency jumps or slope inversion at the characteristic fields of the fixed system hysteresis loop. This allows to assign each mode to one subsystem in an indisputable manner. The modeling of the free layer FMR (Fig.~\ref{FL}) and the  fixed layer system eigenmodes (Fig.~\ref{CoNi}, \ref{CoNi2}) will provide the anisotropies and the interlayer coupling of the essential layers of the MTJ at all stages of annealing (Tables~\ref{TMR}-\ref{bilan}). 

%
\begin{figure}
\caption{(Color online). Loops of the fixed layers system and eigenexcitations thereof after annealing at 300$^\circ$C [(a)-(b)], 350$^\circ$C [(c)-(d)] and 400$^\circ$C [(e)-(f)]. \textcolor{black}{The blue loop is the experimental data, while the black and gray loops are the results of zero temperature 3-macrospin calculations. The cross symbol is panel (a) is there to evidence a slope in the loop at the approach to saturation. The corresponding orientations of the magnetizations of the three layers of the fixed system are sketched in Fig. 4.} The triplets on top of the eigenmode frequency curves illustrate the modeled precession amplitudes in the FeCoB layer (PL, first component of the triplet), in the thin Co layer (RL, second component) and in the Co/Ni multilayer (HL, third component). In the eigenmode fits, the bold line are for an increasing field scan like used in the experiments, while the dotted lines are for the reverse sweep. The inset in panel (e) is the major loop after 400$^\circ$C annealing. The field is varied between $\pm1$ Tesla.}
\label{CoNi}
\end{figure}

\section{Results}
Let us look in detail at the consequences of annealing. It improves the tunneling magneto-resistance (TMR) and the insulator character (Resistance-Area product, RA) of the tunneling oxide (Table~\ref{TMR}). 
\subsection{Properties of the dual MgO free layers}
The free layer minor loops remain square with no significant change of coercivity (not shown). The FMR frequency vs field curves  [Fig.~\ref{FL}(f)] always exhibit the expected V-like shape; its apex is expected at remanence at a frequency $\frac{\gamma_0}{2\pi} (H_k-M_s)$ where $\gamma_0$ is the gyromagnetic factor. This frequency at remanence yields the effective anisotropy. As clear from Fig.~\ref{FL}(f), the effective anisotropy of our dual MgO free layers is very stable against annealing (Table \ref{FL}). This contrasts with single MgO free layers whose anisotropy usually degrades above 375$^\circ$C of annealing \cite{swerts_beol_2015, devolder_evolution_2016}. Besides, the FMR linewidth $\Delta f$ of our dual MgO free layer does not evolve dramatically upon annealing [Fig.~\ref{FL}(a)-(d)]. Both the Gilbert damping and the samples' inhomogenities contribute to the FMR linewidth. The separation of these two contributions can be done classically by looking at the dependence of the frequency linewidth versus FMR frequency. Despite our modest signal-to-noise ratio, we can ascertain that the linewidth is quasi-linear with the FMR frequency [see Fig.~\ref{FL}(g)]). Writing $\Delta f - \Delta f |_{H=0} = 2 \alpha f_\textrm{FMR}$ allows to extract the Gilbert damping $\alpha_\textrm{FL}$ and a measurement of the inhomogeneity of the effective anisotropy field (Table \ref{FL}). Some increase of the damping parameter occurs upon annealing, however its remains below 0.01, i.e. below the values observed in single MgO free layers \cite{devolder_damping_2013, devolder_time-resolved_2016}. In addition to its contribution to the anisotropy, it seems that the MgO capping also acts as an efficient diffusion barrier that helps to maintain a low damping within the free layer. However a slight increase of the free layer inhomogeneity $\frac{1}{\gamma_0} \Delta f |_{H=0}$ is still unfortunately observed upon annealing (Table \ref{TMR}).

\subsection{Properties of the fixed system}
In comparison to the free layer, the fixed system seems to evolve more substantially upon annealing as attested by the gradual inclination of its hysteresis loop (Fig.~\ref{CoNi}). At 300$^\circ$C a first unexpected fact is that the loop of the fixed system comprises 3 apparent switching events. One may naively conclude that the PL, RL and HL switch successively one by one such that the coupling through TaFeCoB would be vanishingly small; however a closer look at the amplitudes of the switching steps indicate that it cannot be the case. 
Another possible scenario would be that the switching at intermediate field [320 mT, Fig.~\ref{CoNi}(a)] corresponds to the synchronous "rigid" flip of the 3 magnetization layers \{PL, RL, HL\} from a \{$\uparrow \uparrow \downarrow$\} to a \{{$\downarrow \downarrow \uparrow$}\} configuration to align the magnetization of the largest moment layer with the applied field, \textcolor{black}{followed by another flip to the saturated \{{$\uparrow \uparrow \uparrow$}\} configuration}. However, this scenario is too simple and we will se that the collinear alignment of the magnetizations \textcolor{black}{is not always the lowest energy configuration especially near the critical fields of the loops}. Even in the simplest case [Fig.~\ref{CoNi}(a)], \textcolor{black}{the approach to saturation has a  substantial slope (see the x symbol in [Fig.~\ref{CoNi}(a)]), which indicates that the state at this field does not consist in perfect collinear alignment of the magnetizations}. A less intuitive magnetization process has thus to be identified.

Modeling as conducted below \textcolor{black}{and summarized in Fig.~\ref{CoNi2} will indicate that the canted parts of the loops correspond tilted magnetization orientations within the fixed system}.
\textcolor{black}{From the loops, a second} finding is the occurrence of some loss of the HL moment between 300 and 350$^\circ$C annealing \textcolor{black}{(see Table~\ref{bilan})}. This loss of HL moment was found \cite{liu_[co/ni]-cofeb_2016} to originate from the diffusion of Cr from the seed layer into the HL.\\ 
Finally there are also two more subtle points to be noticed. Only two modes are detected out of the three expected from the three magnetic parts of the fixed system. Besides, the frequency versus field slope of the highest frequency detected mode in the medium field region is opposite to that which was observed in conventional SAF in which a weak ferro-coupler was used (see for instance ref. \onlinecite{devolder_evolution_2016}). These two facts indicate that the present coupling though TaFeCoB is probably \textcolor{black}{much stronger than} that observed in the earlier Ta spacers \cite{devolder_evolution_2016}.

To get the quantitative properties of each part of the fixed layers system, we have fitted the observed mode frequencies and hysteresis loops to that of 3 coupled macrospins (PL, RL and HL) of unknown anisotropies and couplings. The ground state and its eigenexcitations were calculated by staying in the energy minimum during a field sweep and then by linearizing the dynamics about the energy minima to infer the eigenexcitations. The layer thicknesses were considered constant during annealing such that any interdiffusion process will be accounted for by the layers' magnetizations. The approach used to vary the unknown material parameters and make the simulated eigenmode frequencies fit with the experimental ones is similar to that used in ref. \onlinecite{devolder_evolution_2016}. The deduced material properties for the layers within the fixed system are gathered in Table~\ref{bilan}, \textcolor{black} {and the resulting layer-resolved modeled loops are reported in Fig.~\ref{CoNi2}.} 

\textcolor{black} {It is interesting to note that for the annealing step of 300$^\circ$C the interlayer exchange energy and the anisotropy energies are both very strong on both sides of the Co RL. As a result, when the interactions can not be all satisfied at the approach to saturation, the system prefers to partition its energy in both the exchange degree of freedom and the anisotropy degree of freedom by transiently forming a non-collinear configuration (see the most stable states in Fig.~\ref{CoNi2}(a) between 460 and 600 mT). This happens in a manner very similar to within a Bloch-type domain wall, where the magnetization is fixed in the two domains and prefers to rotate gradually in space instead of abruptly from one domain to the next.}

%
\begin{figure}
\caption{(Color online). \textcolor{black}{Hysteresis loops of the fixed layers system after 300$^\circ$C (a) and 400$^\circ$C (b) annealing as calculated from the material properties of Table~\ref{bilan}. The colored arrows sketch the orientations of the magnetization of the three layers of the fixed system. The corresponding field positions are indicated by the cross symbols when disambiguation is needed. The loops are calculated for an increasing field. In positive fields, the most stable configuration (narrow arrows) can be reached only if a finite energy barrier is thermally overcome by the system. 
In the experiments related of (a), the HL passes from the metastable $\downarrow \uparrow \uparrow$  to the stable configuration $\uparrow \downarrow \downarrow$ at a field position marked by the vertical green line labeled "\textit{exp. Hc}". When further increasing the field, the PL and RL partially switch and then their magnetizatiosn progressively unwind (grey cross) at 500 mT. In the experiments, the metastable state present above 320 mT (right of the green line) is not visited by the system.}}
\label{CoNi2}
\end{figure}

\section{Discussion}

\begin{table*}
  \centering
  \begin{tabular}{c|c|ccccc}
  \hline
Layer  & polarizing layer  & ferro-coupler & reference layer   & antiferro- & hard layers  	\\
Role  &  (PL) & \& texture-breaker &  (RL)  & coupler &  (HL) 	\\
 Composition & MgO/Fe$_{60}$Co$_{20}$B$_{20}$ & TaFeCoB &  Co(6) & Ru &[Co(6)/Ni(3)]$_{\times 6}$ / NiCr \\ 
Thickness ($t$, \r{A})  & 8 &  8 &   6 & 4 & 54 \\ \hline \hline
  &  $M_S$ (A/m)  &$J_{\textrm{TaFeCoB}}(\pm$ 0.03 mJ/m$^2$)  & $M_S$ (A/m)  & $J_{\textrm{Ru}}(\pm$ 0.03 mJ/m$^2$)  & $M_S$ (A/m)\\   
300$^{\circ}$C &   $ 1.1\times10^{6} $ & $\geq 1.0$  & $1.1\times10^{6} $  & -0.85 & $0.55\times10^{6}$ \\  
350$^{\circ}$C &    $1.1\times10^{6}$ & $\geq 1.47$ & $ 0.6\times10^{6} $  & -0.71 & { $0.49\times10^{6}$ } \\ 
375$^{\circ}$C &   $1.1\times10^{6}$ & 1.17 & $ 0.6\times10^{6} $  & -0.61 & {$0.49 \times10^{6}$}  \\
400$^{\circ}$C &   $1.1\times10^{6}$ & 0.8 & $0.6\times10^{6}$  & -0.5 & {$0.49\times10^{6}$ } \\ \hline \hline
   & $ H_{k}$ (A/m)   &   & $ H_{k}$ (A/m) & & $ H_{k}$ (A/m) &\\  
   & $\mu_0(H_{k}-M_S)$ ($\pm$ 25 mT)   &   & $\mu_0(H_{k}-M_S)$ ($\pm$ 50 mT) & & $\mu_0(H_{k}-M_S)$ ($\pm$ 50 mT) &\\ \hline   
300$^{\circ}$C     & $1.45\times 10^6$ & & $1.35\times 10^6$  & & $1.1\times 10^6$ \\  
    &    440  \# & $\xleftrightarrow{\text{interdependent}}$ &  \# 314    & & 690 \\  \hline 
 350$^{\circ}$C   & $1.3\times 10^6$ & & $0.75\times 10^6$  & & $1.05\times 10^6$ \\   
     & 250  \# & $\xleftrightarrow{\text{interdependent}}$ & \# 188   & & 700 \\  \hline 
 375$^{\circ}$C     & $1.2\times 10^6$ & & $0.65\times 10^6$  & & $1.05\times 10^6$ \\    
  &   125  & & 63 & & 700 \\  \hline 
400$^{\circ}$C  & $1.0\times 10^6$ & & $ 0.2\times 10^6$  & & $ 1.05 \times 10^6$ \\   
   &   -125  & & -500  & &  700 \\  \hline
    \end{tabular}
  \caption{Properties of the fixed layers versus annealing temperature. 
 \# indicates that the parameters cannot be independently determined. }
  \label{bilan}
\end{table*}

Several of our findings confirm previous conclusions: just like in MTJs with a conventional SAF, the FeCoB polarizing layer undergoes a gradual and moderate loss of anisotropy upon annealing \cite{devolder_evolution_2016}. Also anticipated, the anisotropy of the [Co/Ni] hard layer is strong and stays practically unaffected by annealing \cite{liu_[co/ni]-cofeb_2016}. However, we can notice two substantial differences compared to the standards set by conventional (i.e. thick) SAF in which a Ta texture breaking layer is used to separate the spin polarizing layer from the reference layer. 

First, the exchange couplings through TaFeCoB is particularly strong and it persists upon annealing. This has two consequences. The first is that the optical excitation of the \{PL + PL\} ensemble is predicted to be pushed above our measurement upper limit of 70 GHz, hence above our detection capability. As a result, we can only give a lower bound of the coupling through the TaFeCoB spacer at the two lowest annealing temperatures. The \{PL +RL\} ensemble behaves like a rigid block and the anisotropies of the PL and RL are somewhat interchangeable (only their weighed sum can be deduced). \\
The second and more visible consequence of the very strong interlayer exchange through both TaFeCoB and Ru is that for some applied fields, the layers' magnetizations cannot satisfy both their anisotropy and their exchange couplings. The Co RL layer often ends in a tilted orientation state, like in a spring magnet. The unanticipated \textcolor{black}{tail in the approach to saturation in the loops} at positive applied field in Fig.~\ref{CoNi}(a-c) is the signature of the gradual unwinding of such vertical wall within the fixed system. \\

Second, the 6\r{A} thick cobalt reference layer seems to be the main weakness of our MTJ: upon annealing, it quickly looses its anisotropy, which manifests as a loop rounding  [Fig.~\ref{CoNi}(e)]. Fortunately the strong interlayer couplings maintain its perfect perpendicular orientation at remanence, such that the MTJ is functional even at the toughest annealing conditions. Although strong, the interlayer exchange coupling through Ru is below the expected value for a Ru thickness matching with the first antiferromagnetic maximum \cite{zhao_perpendicular_2008}. This indicates that a thickness of 6~\r{A} of Co is not sufficient to fully develop the coupling, as we could confirm with thicker Co layers (not shown) that are however less performing when examining the stray field compensation of the fixed system. \\

\section{Conclusion}
In summary, we have studied the evolution of perpendicularly magnetized MTJs upon annealing up to 400$^\circ$C. Our system comprises a dual MgO free layer that is bottom-pinned by a thin SAF fixed system based on a single Co/Ni multilayer seeded by a NiCr buffer. The loops indicate that the layers have their full remanence after all annealing steps despite the evolution towards an ever increasing propensity to canting inside the fixed system, visible at finite applied field. The FMR eigenmodes were modeled to reveal the evolutions of the layers properties. The dual MgO free layer has a robust anisotropy and suffers only a minor increase of its Gilbert damping  upon annealing. The [Co/Ni] multilayer undergoes a decrease of its moment but keeps an anisotropy field as strong as 700 mT during annealing. Conversely, the FeCoB polarizing layer, and to a more dramatic extent the thin Co reference layer loose their anisotropy upon annealing. The coupling through the texture breaker TaFeCoB layer and the Ru antiferro-coupler are fortunately large enough to ensure full perpendicular remanence of all layers of the MTJ. \\
This diagnosis argues however for the search of new material solutions that would avoid the partial loss of anisotropy of the reference layer that might be detrimental to device reliability. Options include for instance the insertion of Pt between Co and Ru; this was proven efficient \cite{bandiera_enhancement_2012} to strengthen the Ru-mediated coupling and the Co anisotropy, yet studied only for thicker Ru spacers so far. Alternative texture-breaking layers \cite{devolder_ferromagnetic_2016-1} that would be more immune to interdiffusion upon annealing may also deserve to be studied as a replacement option to TaFeCoB. 

%

\end{document}